\documentclass[prb,twocolumn,aps,superscriptaddress]{revtex4-1}
\usepackage{graphicx}
\usepackage{dcolumn}
\usepackage{bm}
\usepackage{graphicx}
\usepackage{subfigure}
\usepackage{physics}
\usepackage[english]{babel}
\usepackage[usenames, dvipsnames]{color}
\usepackage{hyperref}

\begin{document}
	
	
	\title{Three terminal vibron coupled hybrid quantum dot thermoelectric refrigeration}
	
	\author{Swarnadip Mukherjee}
	\email{swarnadip@ee.iitb.ac.in}
	\thanks{Equal contribution}
	\author{Bitan De}
	\thanks{Equal contribution}
	\author{Bhaskaran Muralidharan}%
	\email{bm@ee.iitb.ac.in}
	\affiliation{Department of Electrical Engineering, Indian Institute of Technology Bombay, Powai, Mumbai-400076, India
	}
	
	
	\date{\today}
	
	\begin{abstract}
		A three terminal nanoscale refrigeration concept based on a vibron-coupled quantum dot hybrid system coupled to two electronic reservoirs and a phonon bath is proposed and analyzed in detail. While investigating the non-trivial role of electron-phonon interactions, we show that, although they are well known to be detrimental from a general refrigeration perspective, can be engineered to favorably improve the trade-off between the cooling power (CP) and the coefficient-of-performance (COP). Furthermore, an additional improvement in the trade-off can be facilitated by applying a high electronic thermal bias. However, the allowed maximum of the thermal bias being strongly limited by the electron-phonon coupling, in turn, determines the lowest achievable temperature of the cooled body. It is further demonstrated that such interactions drive a phonon flow between the dot and bath whose direction and magnitude depend on the temperature difference between the dot and bath. To justify its impact in optimizing the peak CP and COP, we show that a weak coupling with the bath is preferable when the phonons relax through it and a strong coupling is suitable in the opposite case when the phonons are extracted from the bath. Finally, in studying the effect of asymmetry in electronic couplings, we show that a stronger coupling is favorable with the contact whose temperature is closer to that of the bath. Combining these aspects, we believe that this study could offer important guidelines for a possible realization of molecular and quantum dot thermoelectric refrigerator.

	\end{abstract}
	
	\maketitle
	\section{Introduction}
\begin{figure}[!htbp]
		\centering
		\includegraphics[height=0.38\textwidth,width=0.45\textwidth]{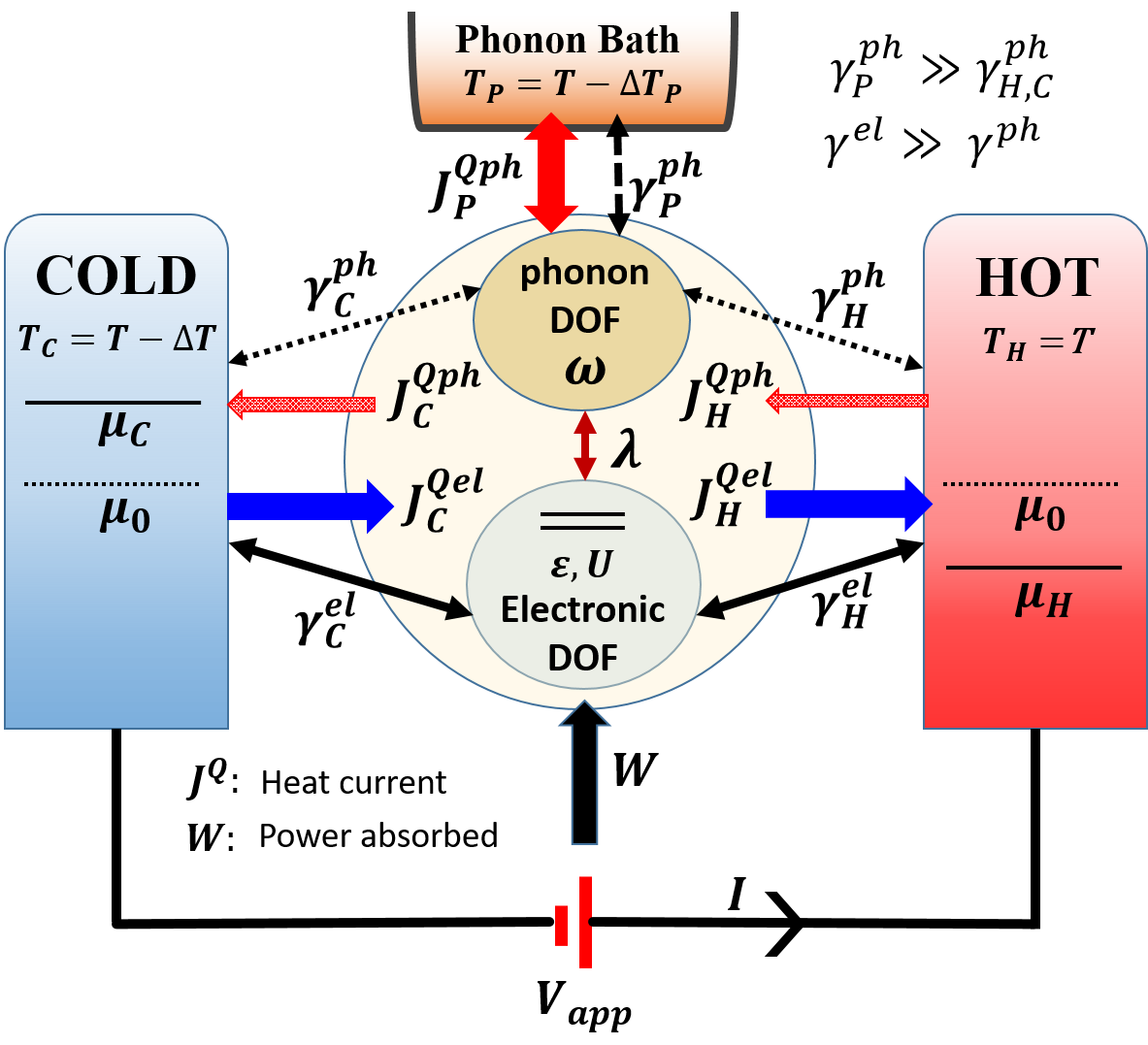}
		\quad
		\caption{Setup schematic and model: Schematic of a three-terminal refrigerator setup which comprises a vibron-coupled QD connected to two electronic reservoirs having different temperatures and a phonon bath, that captures the heat exchange between the dot and substrate. The dot constitutes both electronic and phonon degrees-of-freedom which are coupled by the electron-phonon interaction ($\lambda$). A voltage bias applied between the contacts facilitates the flow of charge and heat currents from the cold to hot terminal. Refrigeration performance is governed by the coupled transport of charge and heat currents through the entire setup.}
		\label{dev}
	\end{figure}

	Exploring new vistas on nanoscale thermoelectric power generation, i.e., the conversion of waste heat into electrical energy, has in the past two and a half decades seen feverish activity both on the theoretical \cite{Dressel1,Mahan,Humphrey2005} and experimental fronts \cite{QDTEexptPRL19,EXPTnatcom15,EXPTnatnano15,RefriEXPTprb14}. A major conceptual breakthrough en route was the proposal on the exploitation of sharp spectral features through nanostructuring \cite{Dressel1,Mahan,Humphrey2005} to achieve a high figure-of-merit in the linear transport regime. Subsequently, studies in the nonlinear transport regime \cite{Hershfield,jiangNL,JiangPRA17NL} have also gained precedence in order to address the power-efficiency trade-off \cite{Muralidharan2012,akshaybm,Naka,Sothmann_Review,De2016,pankaj,myTED,myPRA,myTF}. Here, aspects that include lineshape engineering \cite{myTED,myPRA,myTF,Whitney2014,Whitney2015} and energy filtering \cite{Bahk,Aniket2017} have been explored in great detail. \\
	\indent The complementary aspect, that is, nanoscale refrigeration using similar concepts \cite{EdwardsQD,Edwardscooling}, namely sharp spectral features observed in quantum dot (QD) systems,  was relatively less explored until recently with lot of attention \cite{natnano15,RefriEXPTprb14,CbarrierPRB,chenPRE,Dare,2DEGrefri,chiralPRB,absorption} due to the rising demand of on-chip cooling \cite{natnano15}. Thermoelectric refrigeration involves, at a very basic level, the flow of an electronic charge current accompanied by energy exchange with bosons \cite{CBHPRL,DattaQT} via electron-boson interaction, resulting in a heat current that cools the cold body by heating the hot end. 
	In general, the study of correlated charge and energy transport requires a deeper insight into the many-particle interactions of electrons with ancillary subsystems like phonons \cite{Reddy,Aniket_Peltier}, photon cavity modes \cite{Nori,Kontos}, magnons \cite{Magnonics} and nuclear spins \cite{Levitov,Hirayama,Abhin}, and as such can be categorized within the active research area of hybrid quantum systems. In particular, there have been a lot of experimental advances on hybrid systems featuring the interplay of electronic modes with vibrational modes driven out of equilibrium with the vision of engineering these modes for various quantum technology applications \cite{Leroy,Imran,natnano15,Mandar_1}.\\
	\indent In this paper, we propose and investigate in detail a three terminal nanoscale refrigeration concept \cite{joint} based on a vibron-coupled QD hybrid system  \cite{Ilani,siddique,Arrachea,De2016,De_2019,Desrep}. The setup schematic elucidated in Fig.~\ref{dev}(a) is modeled using a three-terminal geometry which comprises a single mode vibron-coupled quantum dot connected to two electronic contacts, kept at different temperatures and electrochemical potentials, and a bulk phonon bath at an intermediate temperature that exchanges phonons with the dot under non-equilibrium situations. The earlier works related to three-terminal refrigeration \cite{jiangNL,Dare,3TrefriPRE2020} were primarily focused on cooling down the phonon bath by applying bias across the electronic contacts, and the phonons extracted from the bath are utilized in inelastic electron tunnelings between the conducting channels. In contrast, our model features a variant of the three terminal configuration, where the energy harvesting setup closely resembles a two-terminal geometry and the phonon exchange with the substrate/bulk is represented through a third terminal, modeled as a phonon bath which makes the whole setup a three-terminal one. In this sense, we use only a single Coulomb-blockaded conducting channel which is coupled to all three terminals.\\
	\indent The key objective of this study is to explore the non-trivial aspects of the interplay between the electronic and vibron states and its implications on the refrigeration performance especially when the phonon population in the dot is driven out of equilibrium with respect to the bath. In due course, we note that as the electron-phonon interaction gets stronger, the phonon assisted tunneling (PAT) processes dominate over direct tunneling (DT) resulting an accumulation or absorption of phonons into the dot. The degree of phonon population in the dot and the temperature of the bath resolve the sense of dot-to-bath phonon flow, making the bath either a dissipator or injector of phonons \cite{joint,Mazza_2014}. At a primary glance, PAT leads to a steady fall in the cooling power and refrigeration efficiency accompanied by a shrinkage of the refrigeration window. However, these processes can be tuned to favorably improve the trade-off between the CP and COP. In subsequent illustrations, it is demonstrated that a high electronic thermal bias, jointly with the modulation of electron-vibron coupling, has a vital role to play in improving the trade-off further. However, the allowed maximum of the applied thermal bias being strongly limited by the coupling strength, restricts further improvement of the trade-off and thereby determining the lowest achievable temperature of the cold contact.\\ 
	\indent The next course of this study explores the roles played by the phonon bath and the junction coupling rates in optimizing the peak refrigeration performance. The dot-to-bath phonon flow exhibits a dual nature when the bath temperature is swept across a temperature scale bounded by the cold and hot end. The dot dissipates phonons through the bath as the temperature of the bath approaches towards its minima and reverses its role in the opposite limit. Utilizing the principal of heat balance, we testify that in the former case one should have a strong coupling with the phonon bath to boost the peak refrigeration performance, whereas, this picture becomes exactly opposite in the latter case. Lastly, we reiterate this study in a similar setup by incorporating an asymmetry in the  electronic coupling between the dot and the contacts considering its relevance in the system level from fabrication point of view. We note that a stronger coupling with the cold contact is favorable when the temperature of the bath is close to the cold end, while a strong coupling with the hot contact is desired in the opposite case.\\
	\indent The rest of the paper is organized as follows. In Sec.~\ref{physics}, we elaborate the model Hamiltonian, transport formalism and discuss key refrigeration parameters. In Sec.~\ref{neqph}, we analyze the refrigeration performance via electron-phonon coupled transport. The trade-off characteristics and optimization study are presented in Secs.~\ref{trdoff} and \ref{opti}, respectively. We conclude in Sec.~\ref{conclu}.
	\section{Physics and Formulation}
	\label{physics}
	\indent A schematic of the three-terminal device studied here is presented in Fig.~\ref{dev}. It comprises a vibron-coupled QD as the central part coupled to three macroscopic reservoirs (or terminals), namely two electronic contacts ($H$ and $C$) and one thermal bath ($P$).  A charge current $I$, is driven by the applied voltage ($V_{app}$) or thermal gradient $\Delta T=(T_H-T_C)$, or both. We will now describe the formulation and working principles of this set up and its refrigeration functionality, which is governed by the coupled transport of charge and heat currents through the entire setup.\\
	\begin{figure}[!htbp]
		\centering
		\includegraphics[height=0.45\textwidth,width=0.482\textwidth]{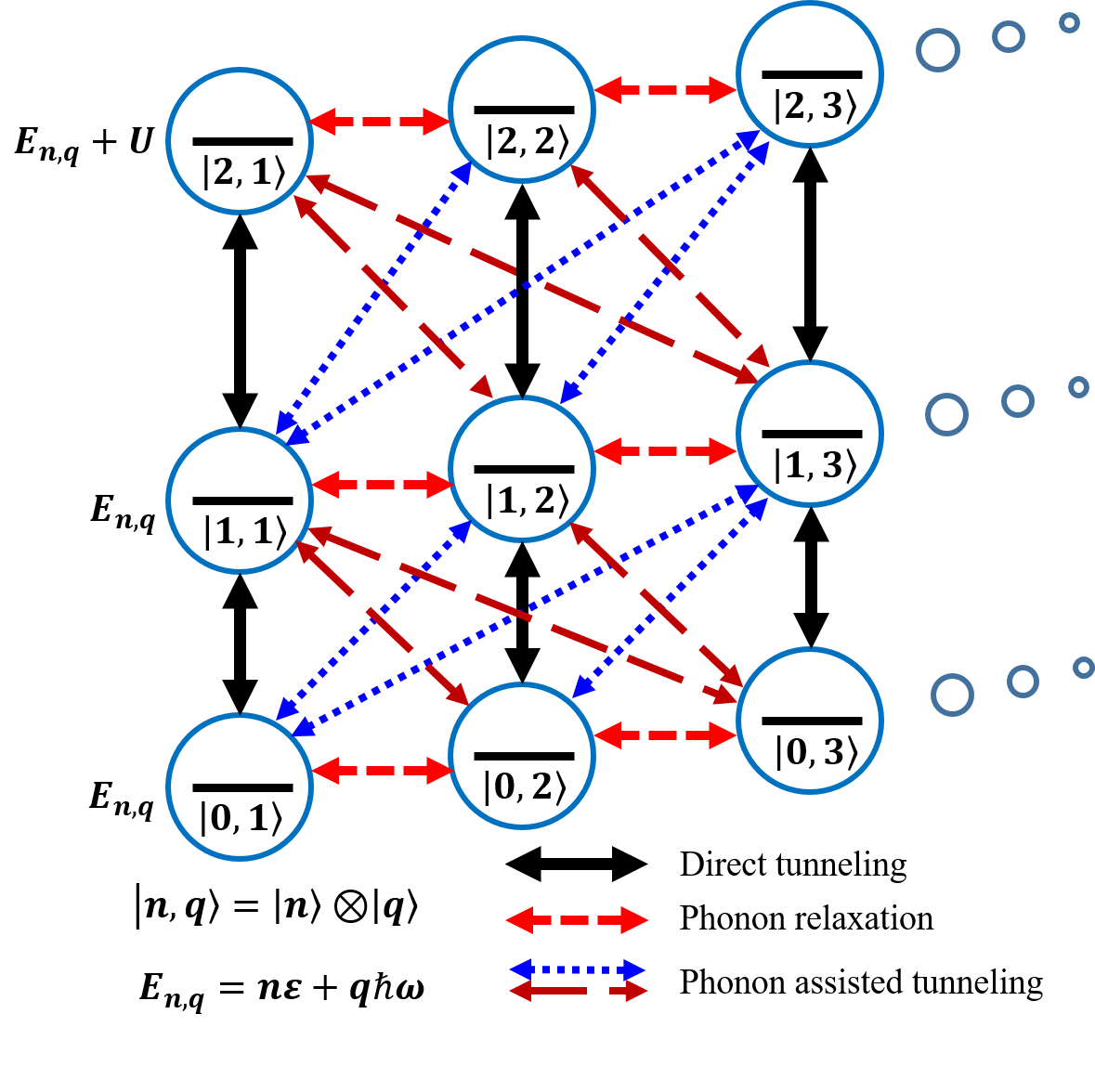}
		\caption{Truncated view of the state transition diagram in the electron-phonon Fock space following the \text{polaron} transformation. Each circle represents an electron-phonon product state $|n,q\rangle=|n\rangle \otimes|q\rangle$ designated by electron number $n$ and phonon number $q$ with energy $E_{n,q}$. Black solid arrows represent direct electron tunneling. Blue and brown dotted arrows signify phonon-assisted electron tunnelings causing phonon emissions ($|n,q\rangle\rightarrow|n+1,q'\rangle$) and absorptions ($|n,q\rangle\rightarrow|n-1,q'\rangle$), respectively. Red dotted arrows stand for bath-assisted phonon transitions ($|n,q\rangle\rightarrow|n,q\pm 1\rangle$). }
		\label{fock}
	\end{figure}
	\subsection{Model Hamiltonian}
	\indent The composite Hamiltonian of the three terminal setup is expressed as $\hat{H}=\hat{H}_D+\hat{H}_R+\hat{H}_P+\hat{H}_{DR}+\hat{H}_{DP}$, where $\hat{H}_D$, $\hat{H}_R$ and $\hat{H}_P$ represent the Hamiltonians of the quantum dot, electronic reservoirs or contacts ($H$ and $C$) and external thermal bath ($P$), respectively. On the other hand, $\hat{H}_{DR}$ and $\hat{H}_{DP}$ describe the dot-to-contact electron tunnelings and dot-to-bath phonon relaxation processes, respectively. The dot Hamiltonian $\hat{H}_D$ is further subdivided as 
	\begin{equation}
		\hat{H}_D=\hat{H}_{el}+\hat{H}_{ph}+\hat{H}_{el-ph},
	\end{equation} 
	where $\hat{H}_{el}$ and $\hat{H}_{ph}$ stand for the electronic and vibron degrees of freedom within the dot,  and $\hat{H}_{el-ph}$ represents the electron-phonon coupling Hamiltonian between the electronic level and the single vibron mode. The electronic part $\hat{H}_{el}$ consists of a spin degenerate energy level with an on-site energy $\epsilon$ and a finite Coulomb interaction energy $U$ for double occupancy. The phonon part $\hat{H}_{ph}$ consists of a single vibron mode $\nu$ (vibron) with angular frequency $\omega_{\nu}$ and the interaction Hamiltonian $\hat{H}_{el-ph}$ is a function of dimensionless coupling strength of the electron-phonon interaction $\lambda_{\nu}$ within the dot. The Hamiltonians $\hat{H}_{el}$, $\hat{H}_{ph}$ and $\hat{H}_{el-ph}$ are written as
	\begin{equation}
		\begin{split}
			\hat{H}_{el}=\sum_{\sigma\in \uparrow,\downarrow}^{}\epsilon d_{\sigma }^{\dagger}d_{\sigma}+Ud_{\uparrow}^{\dagger}d_{\uparrow}d_{\downarrow}^{\dagger}d_{\downarrow},\\ 
			\hat{H}_{ph}=\hbar\omega_{\nu}b_{\nu}^{\dagger}b_{\nu},\\
			\hat{H}_{el-ph}=\sum_{\sigma}^{}\lambda_{\nu}\hbar\omega_{\nu}(b_{\nu}^{\dagger}+b_{\nu}) d_{\sigma }^{\dagger}d_{\sigma},
		\end{split}
	\end{equation}
	where $d_{\sigma}^{\dagger}$($d_{\sigma}$) are the creation (annihilation) operators for the electrons with spin $\sigma$ and $b_{\nu}^{\dagger}$($b_{\nu}$) represent the phonon creation (annihilation) operators of the vibron mode $\omega_{\nu}$. The QD is weakly coupled with the macroscopic reservoirs. Among the reservoirs, the electronic contacts ($H$ and $C$) constitute non-interacting electrons with momentum states labeled by $k$ and spin $\sigma$. On the other hand, the thermal bath consists of numerous non-interacting phonon modes $r$. The Hamiltonians $\hat{H}_R$ and $\hat{H}_P$ read
	\begin{equation}
		\begin{split}
			\hat{H}_R=\sum_{(R\in H,C),k,\sigma}^{}\epsilon_{kR\sigma}\hat{c}_{kR\sigma}^{\dagger}\hat{c}_{kR\sigma},\\
			\hat{H}_P=\sum_{r}^{}\hbar \omega_{r}\hat{B}_{r}^{\dagger}\hat{B}_{r}.
		\end{split}
	\end{equation}
	Here, $\hat{c}_{kR\sigma}^{\dagger}$($\hat{c}_{kR\sigma}$) creates (annihilates) an electron in contact $R\in H,C$ with momentum $k$ and spin $\sigma$ with energy $\epsilon_{kR\sigma}$. Similarly, $B_{r}^{\dagger}$($B_{r}$) creates (annihilates) a phonon of angular frequency $\omega_{r}$ in the thermal bath. If the electrons in the contacts are coupled to the electrons in the dot by an energy $\tau_{k,\sigma}^{el}$ and the phonons in the bath are coupled to the phonons in the dot by an energy $\tau_{r,\nu}^{ph}$, then the coupling Hamiltonians are put down as
	\begin{equation}
		\begin{split}
			\hat{H}_{DR}=\sum_{(R\in H,C),k,\sigma}^{}\bigg[\tau_{k,\sigma}c_{kR\sigma}^{\dagger}d_{\sigma}+h.c\bigg],\\
			\hat{H}_{DP}=\sum_{r}^{}\bigg[\tau_{r,\nu}(B_{r}^{\dagger}+B_{r})(b_{\nu}^{\dagger}+b_{\nu})\bigg],
		\end{split}
	\end{equation}
	where $h.c$ stands for the hermitian conjugate. Throughout our work, we assume that the electron tunneling processes conserve the momentum and spin and the phonon transitions are mode independent. These tunneling coefficients are respectively represented as $\tau^{el}$ and $\tau^{ph}$.\\
	\indent Next, we need to diagonalize the dot Hamiltonian $H_D$ to remove the linear term $H_{el-ph}$. To execute this, we employ polaron transformation on the dot Hamiltonian (such that, $\hat{\tilde{H_D}}\rightarrow e^{\hat{S}} H_{D}e^{-\hat{S}}$, where $\hat{S}=\sum_{\nu}^{}\lambda_{\nu}[b_{\nu}^{\dagger}-b_{\nu}]$ is the transformation operator \cite{MahanMP}). Following this, the on-site energy and Coulomb interaction energy are renormalized as
	\begin{equation}
		\begin{split}
			\tilde{\epsilon}=\epsilon-\lambda_{\nu}^2\hbar\omega_{\nu},\\
			\tilde{U}=U-2\lambda_{\nu}^2\hbar\omega_{\nu}.
		\end{split}
	\end{equation}
	After the diagonalization, the eigenstates of $\tilde{\hat{H}}_{D}$ are expressed as the electron-phonon product states which are denoted by $|n,q\rangle$, where $n$ and $q$, respectively, designate the electron and phonon numbers of that state. The corresponding eigen-energies are given by  $E_{n,q}=\epsilon_n+q\hbar\omega_{\nu}$, where $\epsilon_{0}=0$, $\epsilon_{1}=\tilde{\epsilon}$ and $\epsilon_{2}=2\tilde{\epsilon}+\tilde{U}$ for $n=0,1,2$. One must note that the polaron transformation modifies the electronic coupling coefficient $\tau^{el}$  to $\tilde{\tau^{el}}=\tau^{el} exp[\lambda_{\nu}(b_{\nu}^{\dagger}-b_{\nu})]$ implying the dressing of electron tunneling energy by the electron-phonon interaction. However, the change of phonon coupling coefficient is ignored considering weak dot-to-bath coupling.\\
	\indent Before delving into the transport formalism, it is customary to define the dot-to-contact electron tunneling rates and dot-to-bath phonon relaxation rates. To do this, we assume $\tilde{H}_{D}$ to be weakly perturbed by the reservoir Hamiltonians which allows us to compute electron and phonon transition rates via the Fermi's golden rule. The rates of electron tunneling $\gamma^{el}_{R}$ and phonon relaxation $\gamma^{ph}_{R'}$ are thus derived as
	\begin{equation}
		\begin{split}
			\gamma_{R}^{el}=\frac{2\pi}{\hbar}\sum_{R\in H,C}^{}|\tilde{\tau}_{el}|^2\rho_{R,\sigma'},\\
			\gamma_{R'}^{ph}=\frac{2\pi}{\hbar}\sum_{R'\in R,P}^{}|\tilde{\tau}_{ph}|^2 D_{R'},
		\end{split}
	\end{equation}
	where $\rho_{R,\sigma'}$ and $D_{R'}$ are the constant electron and phonon density of states associated with the contacts $R\in H,C$ and $R'\in R,P$, respectively. Since the phonon relaxation through the bath is much more stronger than that with the electronic contacts, one can assume $\gamma^{ph}_P>>\gamma_{H,C}^{ph}$. From now on, we shall simply denote $\gamma^{ph}_P$ as $\gamma^{ph}$.
	With the introduction of model Hamiltonians and the derivation of electron and phonon transition rates, we have set a perfect ground to formulate the transport methodology, which will be analyzed in the next subsection. 
	\subsection{Transport formalism}
	We initiate our discussion on the transport methodology by elaborating the key approximations we made. First, the rate of dot-to-bath phonon relaxation processes is set much lower than the rate of dot-to-contact electron tunneling processes ($\gamma^{el}>>\gamma^{ph}$) to exclude the system damping \cite{Braig}. Additionally, in the absence of damping, the phonon currents emitting from different terminals remain uncorrelated and hence they can be computed separately. Next, we formulate all the transport calculations in the sequential tunneling limit \cite{SegalPRL}, where the energies associated with electron tunnelings and phonon relaxations are assumed to be much lesser than the thermal energy ($\hbar\gamma^{el},\hbar\gamma^{ph}<<k_B T$) \cite{Beenakker}. In this limit, quantum transport through a spin degenerate energy level coupled to the macroscopic reservoirs is calculated via rate equations in the diagonal subspace of the reduced density matrix \cite{Muralidharan2006,Muralidharan2013}. Lastly, we ignore the overlap of two consecutive phonon sidebands by setting the energy gap between two sidebands much higher than the tunnel induced broadening of the energy levels ($\hbar\omega_{\nu}>>\hbar\gamma^{el}$). With this assumption, transport equations can be formulated in the realm of Markov approximation and two consecutive electron tunnelings remain completely uncorrelated. This justifies the use of rate equations which discard the off-diagonal coherence terms of the reduced density matrix.\\
	\indent The rate of electron tunneling between the two states $|n,q\rangle$ and $|n\pm 1,q'\rangle$ (refer Fig.~\ref{fock}) depends on the contact Fermi-Dirac function of the energy difference between the two states and is given by
	\begin{equation}
		\begin{split}
			\Gamma_{R, (n,q)\rightarrow (n+1,q')}^{el}=\gamma_R^{el}|\langle n,q|\tilde{\hat{d}}_{\sigma}|n+1,q'\rangle|^2\\
			\times f_R(E_{n+1,q'}-E_{n,q}),\\
			\Gamma_{R, (n,q)\rightarrow (n-1,q')}^{el}=\gamma_R^{el}|\langle n,q|\tilde{\hat{d}}_{\sigma}^{\dagger}|n-1,q'\rangle|^2\\
			\times [1-f_R(E_{n+1,q'}-E_{n,q})],
		\end{split}
	\end{equation} 
	where $f_{R}(\zeta)=1/[1+exp(\frac{\zeta-\mu_{R}}{k_B T_{R}})]$ is the Fermi-Dirac distribution function associated with the contact $R\in H,C$ with chemical potential $\mu_{R}$ and temperature $T_{R}$. On the other hand, the rate of phonon absorption and emission between the the states $|n,q\rangle$ and $|n,q\pm1\rangle$ obey the quasi-equilibrium Boltzmann's ratio and is expressed as
	\begin{equation}
		\begin{split}
			\Gamma_{R', (n,q)\rightarrow (n,q+1)}^{ph}=\gamma^{ph}_{R'}(q+1) exp\bigg[-\frac{\hbar\omega_{\nu}}{T_{R'}}\bigg],\\
			\Gamma_{R', (n,q)\rightarrow (n,q-1)}^{ph}=\gamma_{R'}^{ph}(q+1),
		\end{split}
	\end{equation}
	where $R'\in R,P$. As we plug in the different electronic tunneling and phonon relaxation rates into the rate equation, it takes the following form in the electron-phonon Fock space (shown in Fig.~\ref{fock}) \cite{De2016} :
	\begin{equation}
		\begin{split}
			\frac{dP_{n,q}}{dt}=\sum_{q'=0,(R\in H,C)}^{q'=N_q}\bigg[\Gamma_{R,(n',q')\rightarrow(n,q)}^{el}P_{(n',q')} \\ -\Gamma_{R,(n,q)\rightarrow(n',q')}^{el}P_{(n,q)}\bigg] \times\delta(n\pm 1,n') \\ +\sum_{R'\in R,P}^{} \bigg[\Gamma_{R',(n,q')\rightarrow(n,q)}^{ph}P_{(n,q')}-\Gamma_{R',(n,q)\rightarrow(n,q')}^{ph}P_{(n,q)}\bigg]\\\times\delta(q\pm 1,q').
		\end{split}	
	\end{equation}
	In the steady-state, the derivative in the left hand side of the equation becomes zero and one can calculate many-body electron-phonon probabilities $P_{n,q}$ by solving a set of algebraic equations through numerical computation. After evaluating $P_{n,q}$, we calculate the charge current ($I_{R}$), electronic heat current ($J_{R}^{Qel}$) associated with the contacts $R\in H,C$, and the phonon current ($J_{R'}^{Qph}$) flowing through the reservoirs $R'\in H,C,P$ as follows
	\begin{equation}
		\begin{split}
			I_{R}=\sum_{q=0}^{N_q}\sum_{q'=0}^{N_q}-e\bigg[\Gamma_{R,(n+1,q')\rightarrow(n,q)}^{el}P_{(n+1,q')}-\\\Gamma_{R,(n,q)\rightarrow(n+1,q')}^{el}P_{(n,q)}\bigg],\\
		\end{split}
		\label{I}
	\end{equation}
	\begin{equation}
		\begin{split}
			J_{R}^{Q el}=\sum_{q=0}^{N_q}\sum_{q'=0}^{N_q}(E_{n+1,q'}-E_{n,q}-\mu_{R})\bigg[\Gamma_{R,(n+1,q')\rightarrow(n,q)}^{el}\\P_{(n+1,q')}-\Gamma_{R,(n,q)\rightarrow(n+1,q')}^{el}P_{(n,q)}\bigg],\\
			J_{R'}^{Qph}=\sum_{q=0}^{N_q}\sum_{q'=0}^{N_q}\hbar\omega_{\nu}\bigg[\Gamma_{R',(n,q)\rightarrow(n,q')}^{ph}P_{(n,q)}-\\\Gamma_{R',(n,q')\rightarrow(n,q)}^{ph}P_{(n,q')}\bigg],
		\end{split}
		\label{JQ}
	\end{equation}
	where $e$ is the electronic charge unit and $N_q$ is the number of maximally allowed phonon sidebands. With the derived expressions of charge current and heat currents, we illustrate the performance metrics of our setup working as a thermoelectric refrigerator. From now on, we will denote $\lambda_{\nu}$ and $\omega_{\nu}$ as $\lambda$ and $\omega$, respectively, due to the presence of a single vibron mode. Also, in the rest of our work we will denote $I_H=-I_C=I$ considering the charge conservation in the channel.
	\begin{figure}[!htbp]
		\centering
		\subfigure[]{\includegraphics[height=0.32\textwidth,width=0.42\textwidth]{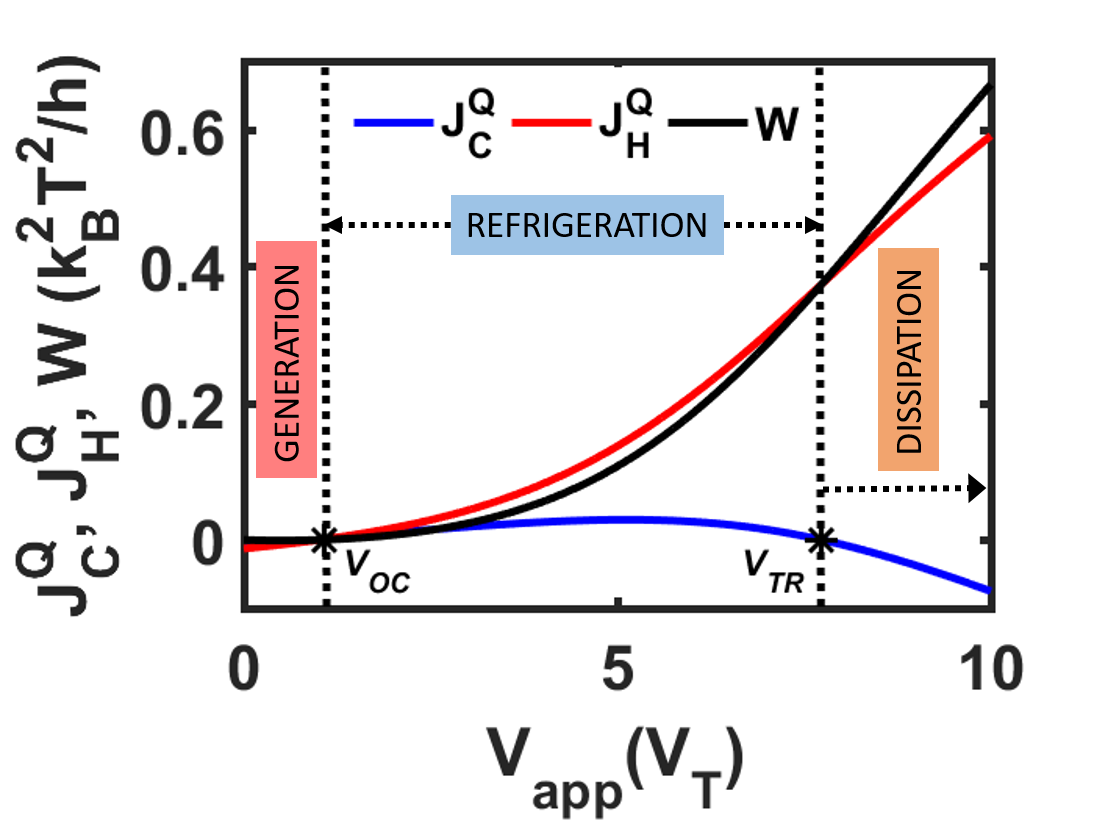}\label{3a}}
		\quad
		\subfigure[]{\includegraphics[height=0.14\textwidth,width=0.4\textwidth]{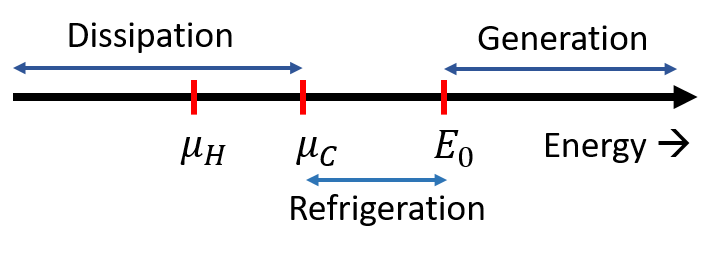}\label{3b}}
		\quad	
		\caption{Refrigeration window: (a) Heat extracted from the cold contact ($J_C^Q$), heat dumped into the hot contact ($J_H^Q$), and the power absorbed from the external source ($W$) are plotted as a function of the applied bias ($V_{app}$). All of them flip signs and become positive beyond the open-circuit voltage ($V_{OC}$) leading to a transition from generation to refrigeration regime. $J_C^Q$ further reverses its direction again at the crossover voltage ($V_{TR}$) and enters into the dissipation regime which restricts the refrigeration in the limit [$V_{OC}$, $V_{TR}$]. (b) Operational regimes and their boundaries for a n-type QD system are shown in an energy scale line plot. Refrigeration occurs when the on-site energy lies in between $E_0$ (reversible energy) and $\mu_C$. Above $E_0$, the device operates in the power generation regime and below $\mu_C$, it enters the dissipation regime.}
		\label{devregime}
	\end{figure}
	\subsection{Refrigeration Performance Parameters} 
	A thermal bias ($\Delta T=T_H-T_C$) applied across the electronic contacts stimulates charge and heat currents. In a typical voltage controlled thermoelectric setup \cite{DattaQT,dattaLNE}, the two contacts are connected by a variable load resistor which causes a potential drop across the contacts. This drop stimulates a back-flow of charge which opposes the flow due to the thermal bias, thereby reducing the net flow. At the open-circuit voltage ($V_{OC}$), these two opposite flows counter-balance each other resulting in a zero current. Therefore, the device works as a power generator in the regime $[0,V_{OC}]$. Beyond $V_{OC}$, the direction of charge flow reverses and the device enters into the power dissipation regime. However, in this regime, if the load is replaced by a voltage source ($V_{app}$) such that power is absorbed into the system, the device functions as a refrigerator as depicted in Fig.~\ref{dev}. In such a scenario, the electronic heat current flows from the cold to hot contact (positive direction as shown in Fig.~\ref{dev} by blue arrows) which effectively cools the cold contact and heats up the hot one. On the other hand, the applied thermal bias sets up a phonon flow from hot to cold contact which opposes the electronic heat flow. Once $V_{app}$ is increased beyond a crossover voltage ($V_{TR}$), the net heat current from the cold contact reverses its direction and the system enters into the dissipation regime as shown in Fig.~\ref{devregime}(a). This restricts the refrigeration in the limit $[V_{OC}, V_{TR}]$ where the dot on-site energy lies in between $E_0$ (reversible energy at which the Fermi-Dirac function of both the contacts are equal) and $\mu_C$ as displayed in Fig.~\ref{devregime}(b).\\
	\indent The system described above mimics a conventional two-terminal refrigerator setup \cite{Muralidharan2012,akshaybm}, where the key performance metrics are the \textit{cooling power} (CP) and the \textit{coefficient-of-performance} (COP). Cooling power is defined as the net heat current extracted from the cold terminal expressed as $J^Q_C=J^{Qel}_C-J^{Qph}_C$ and the COP is defined as
	\begin{equation}
		\eta=\frac{J^Q_C}{W}=\frac{J^{Qel}_C-J^{Qph}_C}{W},
		\label{copex}
	\end{equation}
	where $W$ is the power absorbed from the external source. Utilizing the laws of energy conservation, one can express $W$ in terms of the terminal heat currents as
	\begin{equation}
		W=J^Q_H-J^Q_C,
		\label{hb1}
	\end{equation}
	where $J_H^Q=J^{Qel}_H-J^{Qph}_H$. Since $\gamma^{el}>>\gamma^{ph}$, $J^{Qph}_{H/C}$ becomes negligible as compared to $J^{Qel}_{H/C}$ and hence can be ignored. This approximates the expression of CP and COP as $J^Q_C\approx J^{Qel}_C$ and $\eta\approx\frac{J^{Qel}_C}{W}$, respectively. Following the principle of thermodynamics and using Eq. \eqref{hb1}, one defines the reversible limit of $\eta$ in a two-terminal architecture as $\eta^{2T}_{rev}=T_C/(T_H-T_C)$. Fundamentally, cooling efficiency in a multi-terminal setup is defined as the ratio of the heat current extracted from the cold terminal to the total invested work, both electrical and thermal. The phonon bath considered in this study, being kept at an intermediate temperature in the range $T_C < T_P < T_H$, depending on its dual role as mentioned earlier, defines the refrigeration efficiency in two different ways based on its contribution to the total invested work. 
	In the following, we define the COP and its reversible limit for both the cases considered above. \\
	\textit{Dissipating Bath}: When the bath extracts phonon from the system, the energy-balance equation given in Eq. \eqref{hb1} is modified as 
	\begin{equation}
		J^Q_H+J^{Qph}_P=J^Q_C+W.
		\label{heatbalance}
	\end{equation}
	The expression for COP in such a case remains same as in Eq. \eqref{copex}. However, the reversible limit of $\eta$ in this case is given by (see \ref{app1} for details)
	\begin{equation}
		\eta_{rev}=\eta_{rev}^{2T}\left[ 1+ \frac{J^{Qph}_P}{W} \left( {\frac{T_H}{T_P}-1} \right)\right].
		\label{etarev}
	\end{equation}
	One must notice that $\eta_{rev}$ is no longer a constant quantity and is greater than $\eta_{rev}^{2T}$ as $T_H>T_P$ and $J^{Qph}_P$, $W$ are positive quantities.\\
	\textit{Injecting Bath}: In the opposite limit, when the bath injects phonon into the system and contributes to the total invested work, the COP is defined as
	\begin{equation}
		\eta=\frac{J^Q_C}{W+J^{Qph}_P}.
		\label{copex1}
	\end{equation}
	The energy-balance equation in this case also remains same as in Eq. \eqref{heatbalance} except for a reversal in the sign of $J^{Qph}_P$ which leads to the modified expression of $\eta_{rev}$, given by
	\begin{equation}
		\eta_{rev}=\eta_{rev}^{2T}\left[ 1- \frac{T_H}{T_P} \left( \frac{1}{\frac{W}{J^{Qph}_P}+1}\right)\right].
		\label{etarev}
	\end{equation}
	In contrast to the earlier case, $\eta_{rev}$ becomes less than $\eta_{rev}^{2T}$ which suggests that investing thermal work in a system through a bath, leads to a lowering of the upper bound of the COP. For the rest of this paper, we express the normalized COP as $\eta_r=\frac{\eta}{\eta_{rev}}$.
	
	\section{Results and Discussion}
	\label{results}
	\subsection{Refrigeration via electron-phonon coupled transport}
	\label{neqph}
	\indent In the current section, we elaborate on the impact of vibron-coupled electronic transport in controlling the refrigeration performance. Before going into the details, it is imperative to mention the values of several parameters used in our simulation framework. First, we assume the QD system to be n-type ($\epsilon-\mu>0$), with the dot on-site energy tuned at $3.86$ $k_BT$ above the equilibrium chemical potential ($\mu$) of the contacts, where $T$ is the temperature. Second, the dot-to-contact electronic tunneling rates ($\hbar\gamma_{H,C}^{el}$) are varied in the range $0.015-0.02$ $k_{B}T$, so that the transport physics can be formulated in the realm of sequential tunneling limit. We also ignore the energy dependency of $\gamma_{H,C}^{el}$ considering the wideband approximation. Third, the dot-to-bath phonon relaxation rate ($\gamma^{ph}$) is assumed to be much smaller than $\gamma_{H,C}^{el}$ (typically $\gamma^{ph}\approx 0.1\gamma_{H,C}^{el}$) to exclude damping due to phonon relaxation. Next, the frequency of the vibron mode ($\hbar\omega$) is adjusted to be $0.25$ $k_BT$ in order to validate $\hbar\omega>>\hbar\gamma_{H,C}^{el}$, which justifies the Markov approximation and neglects the effect of bath memory. In molecules or semiconducting quantum dots, the vibrational modes ($\hbar \omega$) usually lie in the range of $(0.1-1)k_BT$ \cite{leijnse,thermalAPL}. Therefore, our assumption of $\hbar \omega=0.25$ $k_BT$ retains a strong analogy with experiments. Lastly, we assume a symmetric capacitive coupling between the dots and contacts so that for a given bias $V_{app}$, the Fermi level of the hot (cold) contact is shifted by $eV_{app}/2$ ($-eV_{app}/2$).\\
	\begin{figure}[!htb]
		\centering
		\subfigure[]{\includegraphics[height=0.225\textwidth,width=0.225\textwidth]{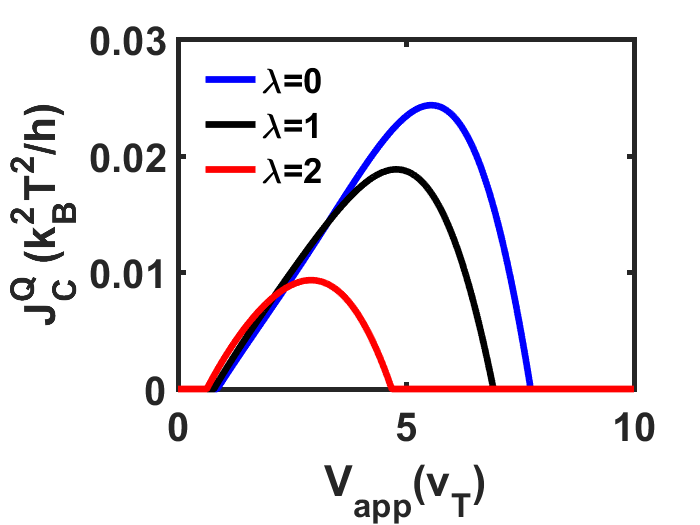}\label{4a}}
		\quad
		\subfigure[]{\includegraphics[height=0.225\textwidth,width=0.225\textwidth]{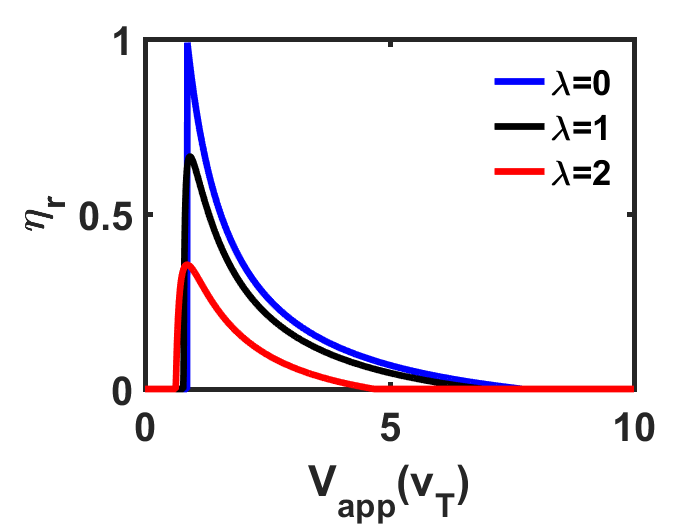}\label{4b}}
		\quad
		\caption{Effect of vibron coupling on refrigeration performance: (a) Variation of cooling power ($J_C^Q$) and (b) normalized COP ($\eta_r$) as a function of $V_{app}$ and $\lambda$. The peak value of $J_C^Q$ ($J_{C,max}^Q$) and $\eta_r$ ($\eta_{r,max}$) exhibit a steady and non-linear fall as $\lambda$ goes up. The voltage corresponding to $J_{C,max}^Q$ and $V_{TR}$ shift toward $V_{OC}$ (which remains almost unchanged) with increasing $\lambda$, resulting in a shrinkage of the refrigeration window.}
		\label{cpcop}
	\end{figure}
	\indent We initiate the discussion of our key results by studying the variation of CP and COP with respect to the applied voltage (normalized by $v_T=k_BT/e$) for different values of $\lambda$. Throughout the paper, a fixed temperature bias across the electronic contacts ($\Delta T=0.2T$) is maintained, unless otherwise specified. Figures \ref{cpcop}(a-b) show that $V_{OC}$ remains almost unaltered with $\lambda$ since $\epsilon-\mu$ is tuned much larger than the thermal energy. 
	However, $V_{TR}$ falls short as $\lambda$ goes up. Referring Eq. \eqref{JQ}, we note that $J_C^{Qel}$ is derived by multiplying each component of $I_C$ by a weighing factor of $(E_{n+1,q'}-E_{n,q})/e$, which is proportional to $\tilde{\epsilon}\equiv\epsilon-\lambda^2\hbar\omega$. It is evident that as $\lambda$ goes up, the weighing factors decrease and leads to the fall of $J_C^{Qel}$. For the same reason, $J_C^{Qel}$ can be brought down to zero at lower $V_{TR}$ with the rise of $\lambda$. In parallel, we notice that the maxima of CP ($J_{C,max}^Q$) registers a sharp fall with $\lambda$ as shown in Fig.~\ref{cpcop}(a). Figure \ref{cpcop}(b) shows the similar falling trend of peak COP ($\eta_{r,max}$) with $\lambda$.\\
	\begin{figure}[!htbp]
		\centering
		\subfigure[]{\includegraphics[height=0.225\textwidth,width=0.225\textwidth]{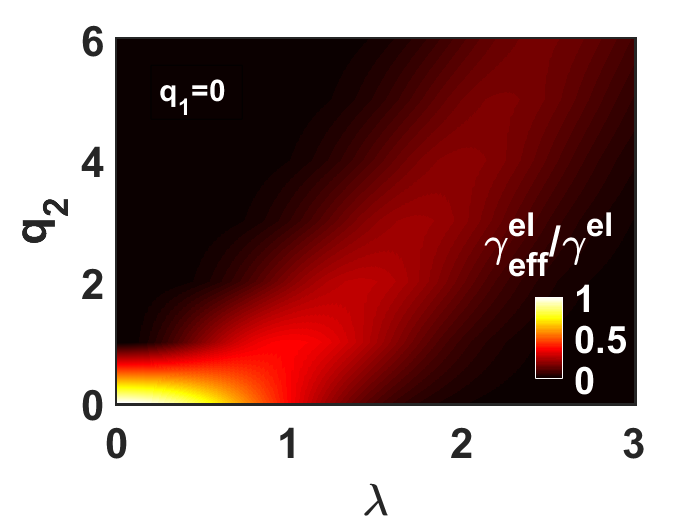}\label{5a}}
		\quad
		\subfigure[]{\includegraphics[height=0.225\textwidth,width=0.225\textwidth]{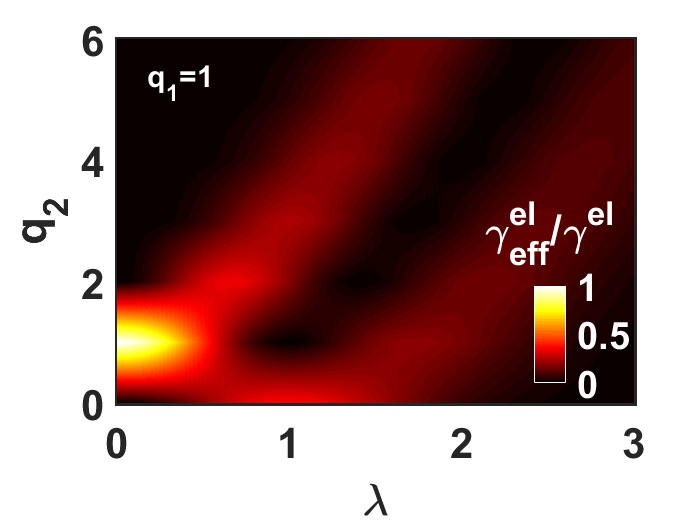}\label{5b}}
		\quad
		\subfigure[]{\includegraphics[height=0.25\textwidth,width=0.25\textwidth]{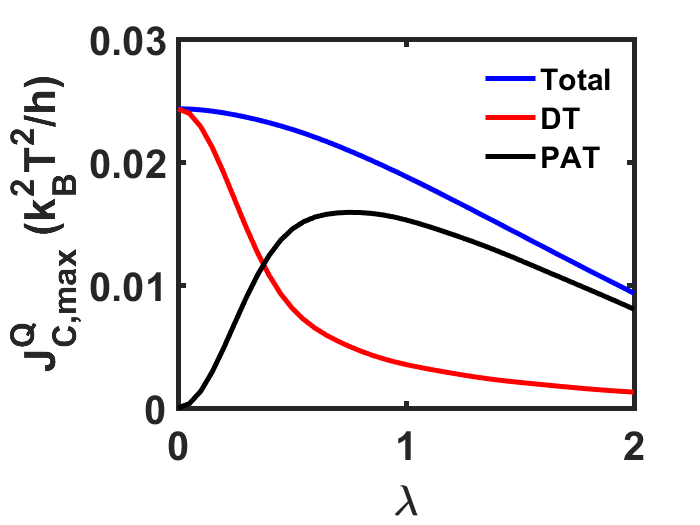}\label{5c}}
		\quad
		\caption{Comparative analysis of DT and PAT: Color map of effective electron tunneling rate ($\gamma_{eff}^{el}/\gamma^{el}$) between two Fock-states with phonon number $q_1$ and $q_2$, as a function of $\lambda$ and $q_2$ for (a) $q_1=0$ and (b) $q_1=1$. Bright spots indicate the DT events ($q_2=q_1$) which dominate over the PAT ($q_2\neq q_1$) at lower values of $\lambda$. (c) Contribution of DT and PAT to total $J_{C,max}^Q$ as a function of $\lambda$. At smaller values of $\lambda$, $J_C^Q$ is dominated by DT, but as $\lambda$ goes up, PAT takes over and primarily controls $J_C^Q$.}
		\label{pat}
	\end{figure}
	\indent Our quantitative analysis estimates $J_{C,max}^Q$ and $\eta_{r,max}$ to be $0.0243$ $k_B^2T^2/h$ and $0.99$, respectively, when electron-phonon coupling vanishes ($\lambda=0$). From the system-design viewpoint, this device when made of a self-assembled array of spherical quantum dots with an average diameter of $40$ $nm$ with a $50\%$ fill factor, can produce a CP density of $0.25$ $MW$/$m^2$ at room temperature. These figures are substantial and can be optimized further by modulating the dot size and filling factors \cite{thermalAPL}. However, in a practical system, these numbers largely degrade due to the presence of electron-phonon coupling which calls for an investigation on the impact of $\lambda$ on CP and COP. From the discussion of the preceding section, we know that a finite $\lambda$ facilitates phonon-assisted-tunneling (PAT) over direct tunneling (DT). Therefore, our next objective is to perform a comparative study on the respective contributions of DT and PAT in $J_{C,max}^Q$.\\
	\indent A finite electron-phonon interaction shifts the potential profile of the dot and modifies the electron-tunneling rate between two Fock states $|n,q_1\rangle$ and $|n\pm 1,q_2\rangle$. The effective tunneling rate ($\gamma_{eff}^{el}$) is then defined as \cite{KochPRL}
	\begin{center}
		\begin{equation}
			\begin{split}
				\gamma_{eff}^{el}=\gamma^{el}\times|FC_{q_1,q_2}|^2
				=\gamma^{el}\times exp\bigg(-\lambda^2\bigg)\times\bigg(\frac{q!}{Q!}\bigg)^2\\\times\lambda^{Q-q}\times L_{q}^{Q-q}(\lambda^2)\times [sgn(q_1-q_2)]^{(q_1-q_2)},
			\end{split}
		\end{equation}  
	\end{center}
	where $q=min(q_1,q_2)$, $Q=max(q_1,q_2)$ and $FC_{q_1,q_2}$ is the measure of the \textit{Frank-Condon} overlap between the two electron-phonon states due to the finite $\lambda$. At $\lambda=0$, the tunneling of electrons do not change the phonon count inside the dot since $\gamma_{eff}^{el}$ vanishes for $q_1\neq q_2$ and the dot retains thermal equilibrium with the bath. As a consequence, the dot-to-bath phonon current freezes \cite{De_2019,Desrep} and the setup closely resembles to a conventional two-terminal architecture. This picture becomes more intriguing at non-zero $\lambda$ when the contribution of PAT to the charge and heat flux become notable as compared to DT. Figures \ref{pat}(a) and \ref{pat}(b)  map the color variation of $\gamma_{eff}^{el}$ between two Fock states ($q_1$ and $q_2$) as a function of $\lambda$ and $q_2$ taking $q_1$ constant, at $0$ and $1$, respectively. In both the cases, $\gamma_{eff}^{el}$ vanishes for $q_1\neq q_2$ at $\lambda=0$ suggesting the existence of only DT events. As $\lambda$ increases, the DT events become less probable and $\gamma^{el}_{eff}$ corresponding to the tunnelings for $q_1\neq q_2$ rises steadily indicating a strong signature of PAT.\\
	\indent Next, we describe the relative contributions of DT and PAT in $J_{C,max}^Q$. Referring to Fig.~\ref{pat}(c), we note that as $\lambda$ increases from zero, the DT component falls drastically and PAT rises steadily, leading to a gradual fall of $J_{C,max}^Q$. With further increase of $\lambda$, PAT component reaches its maxima and starts rolling down thereafter. Meanwhile, the fall of DT component becomes sluggish and tends towards zero for large values of $\lambda$. Therefore, at large $\lambda$, $J_{C,max}^Q$ is solely driven by PAT. As we go up in $\lambda$, Fig.~\ref{pat}(a) shows that for $q_1=0$, $\gamma_{eff}^{el(0,q_2)}$ becomes significant for higher values of $q_2$ which points towards the contribution from higher order side-bands. Figure \ref{pat}(b) also depicts similar pattern for $q_1=1$ except for an additional bright patch which indicates both the up and down PAT transition in the electron-phonon Fock state ladder. With increasing $q_1$, the transition to higher order side-bands with minimal probability becomes significant at higher $\lambda$ resulting in a lowering of current flow. The cumulative effect of all $\gamma_{eff}^{el(q_1,q_2)}$, multiplied with the probability of the respective states $|n,q\rangle$, leads to the signatures of PAT obtained in Figs.~\ref{pat}(a-b).\\
	\indent From our discussion so far, one may apparently label $\lambda$ as a detrimental factor since it brings down $J_{C,max}^Q$, $\eta_{r,max}$ and shrinks the regime of refrigeration $[V_{OC},V_{TR}]$. However, it must be noted that there is a clear improvement in terms of $J_{C,max}^Q-\eta_{r,max}$ trade-off as $\lambda$ increases. Turning back to Fig.~\ref{cpcop}, it is evident that the gap between the operating voltages corresponding to $\eta_{r,max}$ and $J_{C,max}^Q$ becomes narrower with $\lambda$ going up, signaling an improved trade-off between them. In the next subsection, we aim to elaborate this trade-off behavior with respect to different parameters.
	\subsection{Interaction and temperature regulated trade-off between CP and COP}
	\label{trdoff}
	\begin{figure}[!htb]
		\centering
		\subfigure[]{\includegraphics[height=0.225\textwidth,width=0.225\textwidth]{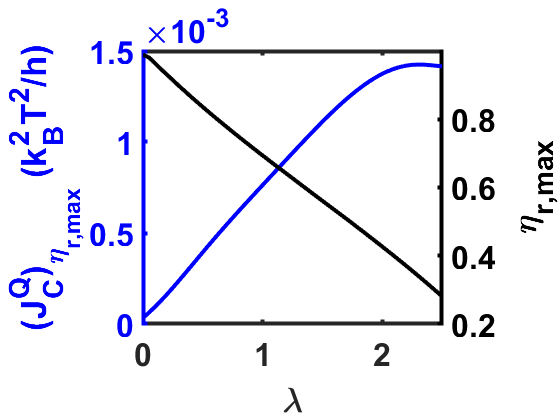}\label{7a}}
		\quad
		\subfigure[]{\includegraphics[height=0.225\textwidth,width=0.225\textwidth]{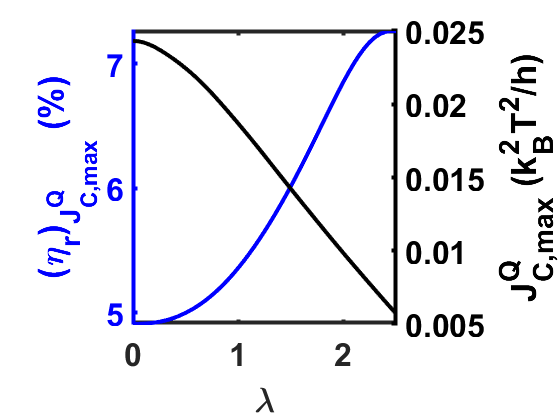}\label{7b}}
		\quad
		\caption{Performance analysis of $(J_{C}^Q)_{\eta_{r,max}}$ and $(\eta_r)_{J_{C,max}^Q}$ with $\lambda$: (a) Double y-ordinate plot of $\eta_{r,max}$ and $(J_{C}^Q)_{\eta_{r,max}}$ and (b) $J_{C,max}^Q$ and $(\eta_r)_{J_{C,max}^Q}$ with respect to $\lambda$. Both $J_{C,max}^Q$ and $\eta_{r,max}$ fall steadily with increasing $\lambda$ as in Fig.~\ref{cpcop}. However, both $(J_{C}^Q)_{\eta_{r,max}}$ and $(\eta_r)_{J_{Cmax}^Q}$ exhibit a steady rise up to a certain $\lambda$ and falls thereafter. These plots indicate that $\lambda$ can be a key factor to facilitate an improvement in the CP-COP trade-off.}
		\label{trdoffmax}
	\end{figure}
\begin{figure}[!htb]
		\centering
		\subfigure[]{\includegraphics[height=0.225\textwidth,width=0.225\textwidth]{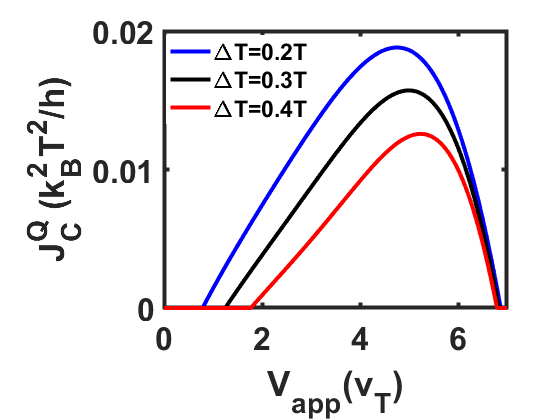}\label{10a}}
		\quad
		\subfigure[]{\includegraphics[height=0.225\textwidth,width=0.225\textwidth]{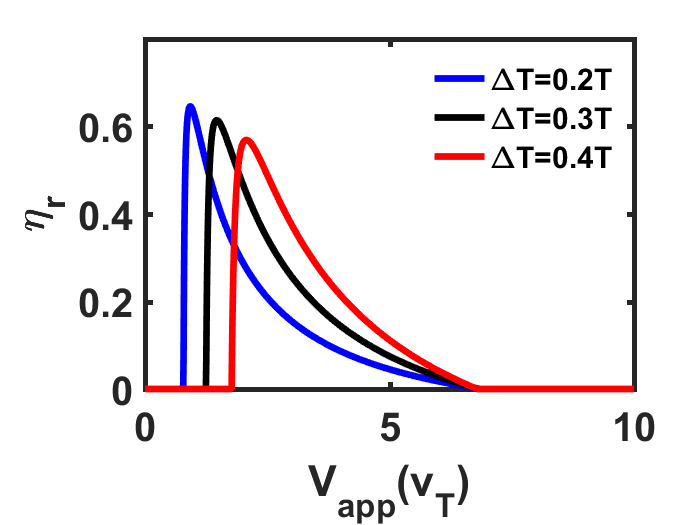}\label{10b}}
		\quad
		\subfigure[]{\includegraphics[height=0.225\textwidth,width=0.225\textwidth]{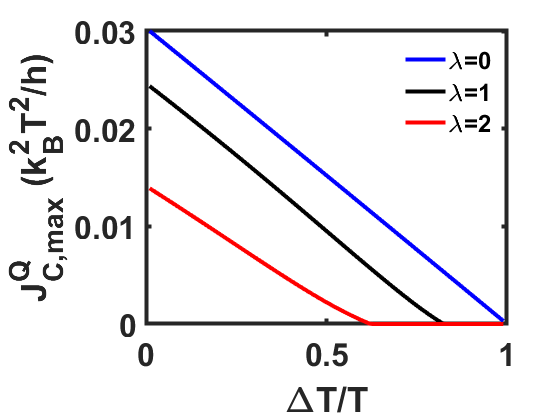}\label{10c}}
		\quad
		\caption{Effect of $\Delta T$ and $\lambda$: Plot of (a) $J_C^Q$ and (b) $\eta_r$ as a function of $V_{app}$ for different $\Delta T/T$. The magnitude of both $J_C^Q$ and $\eta_r$ fall with increasing $\Delta T$ along with a shortening of refrigeration window. Plot of (c) $J_{C,max}^Q$ as a function of $\Delta T$ for different $\lambda$. At large $\lambda$, $J_{C,max}^Q$ approaches zero at a lower $\Delta T$ which limits the lowest achievable value of $T_C$.}
		\label{delT}
	\end{figure}

	\indent We now elaborate on the impact of $\lambda$ in the relative trade-off between CP and COP. Referring to the preceding section (refer to Fig.~\ref{cpcop}), we note that in the absence of $\lambda$, COP maximizes at a voltage where CP is low. Likewise, CP reaches its peak when $\eta_r$ is minimal. Therefore, the optimization of thermoelectric refrigeration calls for the enhancement of two performance metrics, namely (i) the CP extracted at maximum COP denoted as $(J_{C}^Q)_{\eta_{r,max}}$ and (ii) COP acquired at peak CP denoted as $(\eta_r)_{J_{C,max}^Q}$. Here, we test both the quantities by switching on a finite $\lambda$. \\
	\indent Earlier, we note from Fig.~\ref{cpcop} that both $J_{C,max}^Q$ and $\eta_{r,max}$ undergo monotonic decline as $\lambda$ increases. However, the  dependency of $\lambda$ on both $(J_{C}^Q)_{\eta_{r,max}}$ and $(\eta_r)_{J_{C,max}^Q}$ carry some interesting signatures. The double ordinate plot in Fig.~\ref{trdoffmax}(a) exhibits that primarily $(J_{C}^Q)_{\eta_{r,max}}$ increases with $\lambda$ as long as $\lambda$ is small or moderate. As $\lambda$ becomes large, $(J_{C}^Q)_{\eta_{r,max}}$ takes downturn and starts falling steadily. A similar trend is also obtained for $(\eta_r)_{J_{C,max}^Q}$ as depicted in Fig.~\ref{trdoffmax}(b). This observation strongly indicates the significance of the electron-phonon coupling in improving the CP-COP trade-off. In the current existing technology, $\lambda$ can be experimentally tailored in the CNT-based \cite{Ilani,siddique} quantum dots and hence, we believe that this study can lead to reality in terms of optimizing refrigeration efficiency.\\
	\begin{figure}[!htb]
		\centering
		\subfigure[]{\includegraphics[height=0.225\textwidth,width=0.225\textwidth]{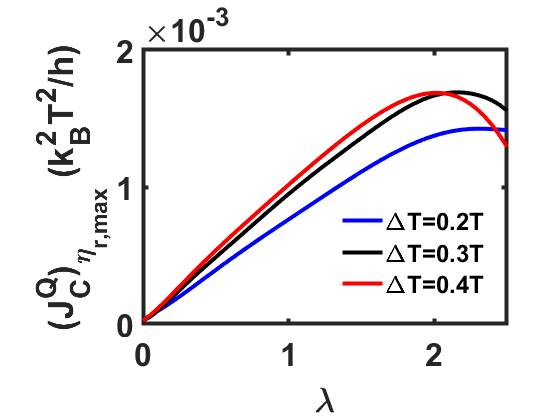}\label{8a}}
		\quad
		\subfigure[]{\includegraphics[height=0.225\textwidth,width=0.225\textwidth]{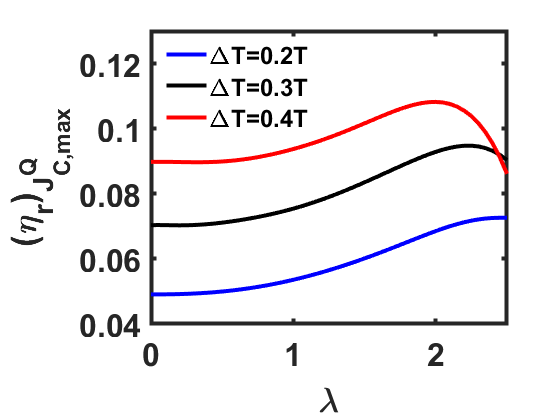}\label{8b}}
		\quad
		\caption{Effect of electronic temperature bias on CP-COP trade-off: Plot of (a) $(J_{C}^Q)_{\eta_{r,max}}$ and (b) $(\eta_r)_{J_{C,max}^Q}$ as a function of $\lambda$ for different values of $\Delta T/T$. As $\Delta T$ increases, both $(J_{C}^Q)_{\eta_{r,max}}$ and $(\eta_r)_{J_{C,max}^Q}$ show a steady enhancement and reach their respective maxima at lower values of $\lambda$. This outcome testifies that electronic thermal bias plays a vital role jointly with $\lambda$ in improving the trade-off.}
		\label{trdoffdelT}
	\end{figure}
	\indent So far, we have performed our simulations by keeping a constant $\Delta T=0.2T$. Now, we re-examine the CP-COP trade-off by varying $\Delta T$. Figures \ref{trdoffdelT}(a) and \ref{trdoffdelT}(b), respectively, note that $(J_{C}^Q)_{\eta_{r,max}}$ and $(\eta_r)_{J_{C,max}^Q}$ increase as $\Delta T$ rises. However, in both the cases we perceive that the maxima of $(J_{C}^Q)_{\eta_{r,max}}$ and $(\eta_r)_{J_{C,max}^Q}$ undergo a non-trivial shift towards the smaller value of $\lambda$ as $\Delta T$ increases. These results necessitate an in-depth analysis on the joint role of $\lambda$ and $\Delta T$ on the non-trivial characteristics of $(J_{C}^Q)_{\eta_{r,max}}$ and $(\eta_r)_{J_{C,max}^Q}$ as addressed below.\\
	\indent First, we inspect the role of applied thermal bias in controlling the allowed refrigeration window and the peak performance. and contribution of $\Delta T$ in the CP-COP trade-off keeping $\lambda$ constant. Figures \ref{delT}(a) and \ref{delT}(b) plot the voltage dependence of $J_{C}^Q$ and $\eta_r$ for different values of $\Delta T$. Drawing an analogy with Figs.~\ref{cpcop}(a-b), we note that both $\lambda$ and $\Delta T$ bring down $J_{C,max}^Q$ and $\eta_{r,max}$ and reduce the refrigeration regime as they go up. However, their prime difference lies in fixing the boundary voltages of the allowed window $[V_{OC},V_{TR}]$. While an increase in $\Delta T$ drags the window towards higher voltage by pushing $V_{OC}$ towards $V_{TR}$, a large $\lambda$ pulls the window towards the low input voltage by lowering $V_{TR}$ and keeping $V_{OC}$ fixed. This has definite consequences on the locus of $J_{C,max}^Q$ and $\eta_{r,max}$ as $\Delta T$ pushes them away while $\lambda$ brings them down. \\
	\indent To explain why the trade-off gets improved jointly by $\Delta T$ and $\lambda$, we look back at Figs.~\ref{cpcop}(a-b) which depict that as $\lambda$ increases, the voltage corresponding to $\eta_{r,max}$ ($V_{\eta_{r,max}}$) moves away from $V_{OC}$. However, the voltage corresponding to $J_{C,max}^Q$ ($V_{J^Q_{C,max}}$) comes closer to $V_{mid}=[V_{OC}+V_{TR}]/2$. As $J_{C,max}^Q$ increases linearly with applied bias around $V_{OC}$ and $\eta_{r,max}$ drops around $V_{mid}$, both $(J_{C}^Q)_{\eta_{r,max}}$ and $(\eta_r)_{J_{C,max}^Q}$ improve up to a moderate value of $\lambda$. Increasing $\Delta T$ also has a similar effect on the trade-off except on $V_{J^Q_{C,max}}$ which moves away from $V_{mid}$ and approaches towards $V_{TR}$.  
	At higher values of $\lambda$ and $\Delta T$, since the magnitudes of $J^Q_C$ and $\eta_r$ fall severely, the trade-off curves reach their maxima and start to fall again. In addition, Fig.~\ref{trdoffdelT} suggests that a large thermal bias leads to the occurrence of trade-off maxima at relatively lower values of $\lambda$.\\
	\indent In connection to the preceding discussion, it should be noted that the trade-off cannot be enhanced indefinitely since the allowed maximum of the applied thermal bias is strongly limited by $\lambda$. Therefore, in this context, one might further be interested in figuring out the lowest achievable limit of $T_C$ at a given $\lambda$ that sustains refrigeration. This can be done by increasing $\Delta T$ until the refrigeration window collapses and $J^Q_{C,max}$ approaches zero. On the other hand, refrigeration requires the CP to be positive ($J^Q_C=J^{Qel}_C-J^{Qph}_C>0$) which imposes a limit on the maximum $\Delta T$ to be applied since $J^{Qel}_C$ falls substantially for large $\Delta T$ while $J^{Qph}_C$ shoots up. The picture becomes more interesting when $\lambda$ is taken into account while calculating minimum $T_C$.\\
	\indent In Fig.~\ref{delT}(c), we plot $J^{Q}_{C,max}$ as a function of $\Delta T$ for different values of $\lambda$. We notice that as $\Delta T$ increases, $J^{Q}_{C,max}$ steadily reduces which is also evident from Fig.~\ref{delT}(a) and at a certain value of $\Delta T$ it reaches zero and the regime of refrigeration ceases to exist. This determines the minimum value of $T_C$ for a given $\lambda$. This feature can be explained with the help of Fig.~\ref{devregime}(b) since the gap between $E_0$ and $\mu_C$ reduces with the reduction of $T_C$. It is also important to note that as $\lambda$ increases, $J^{Q}_{C,max}$ gets substantially lowered and as a result the minimum permissible value of $T_C$ becomes much higher. This suggests that a large electron-phonon interaction strongly limits the range of the allowed thermal bias required for refrigeration.\\
	\begin{figure}[!htb]
		\centering
		\subfigure[]{\includegraphics[height=0.225\textwidth,width=0.225\textwidth]{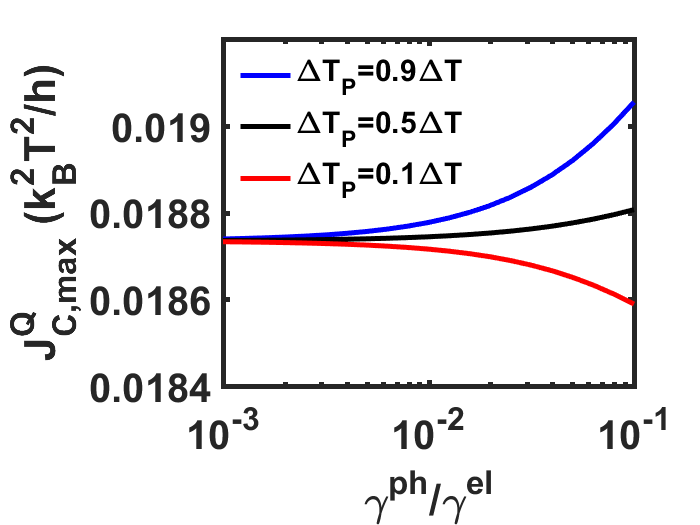}\label{12a}}
		\quad
		\subfigure[]{\includegraphics[height=0.225\textwidth,width=0.225\textwidth]{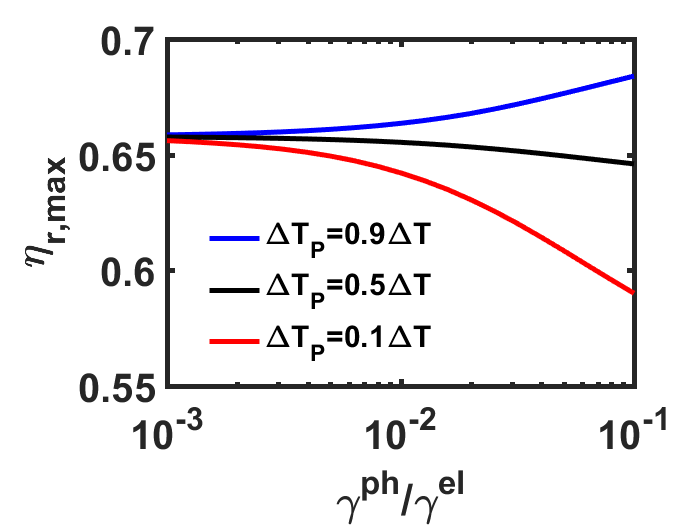}\label{12b}}
		\quad
		\caption{Role of $\gamma^{ph}$: Variation of (a) $J^Q_{C,max}$ and (b) $\eta_{r,max}$ plotted as a function of $\gamma^{ph}$ (normalized by $\gamma^{el}$) at $\lambda=1$ for different values of $T_P$ in semi-log graphs. For $\Delta T_P=0.9\Delta T$, dot dissipates phonons through the bath and $J_P^{Qph}$ increases with $\gamma^{ph}$ causing an enhancement of CP and COP. For $\Delta T_P=0.1\Delta T$, dot absorbs phonons from the bath resulting in a fall of both CP and COP with rising $\gamma^{ph}$. An intermediate situation is observed at $\Delta T_P=0.5\Delta T$.}
		\label{beta}
	\end{figure}
	\indent The discussion thus far highlighted the improvement of trade-off by properly tuning $\lambda$ and $\Delta T$. But, such an improvement is always associated with a steady reduction in the peak performance. In the next section, we discuss the roles played by the phonon bath and the junction coupling rates in optimizing the peak performance.
	\subsection{Performance optimization via bath engineering}
	\label{opti}
\begin{figure}[!htb]
		\centering
		\subfigure[]{\includegraphics[height=0.225\textwidth,width=0.225\textwidth]{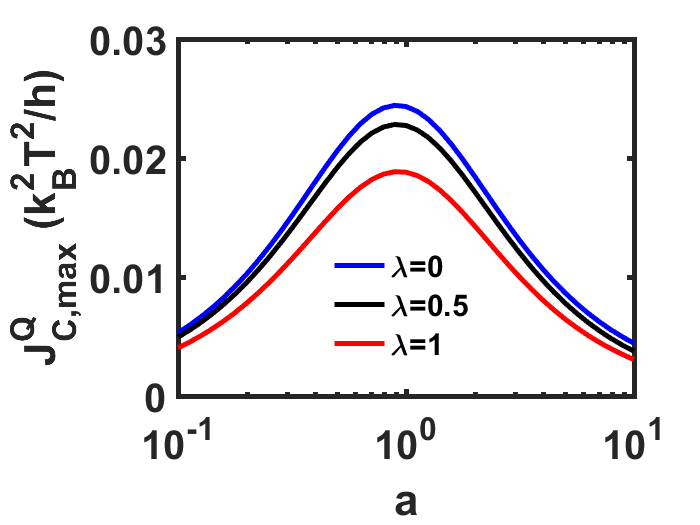}\label{13a}}
		\quad
		\subfigure[]{\includegraphics[height=0.225\textwidth,width=0.225\textwidth]{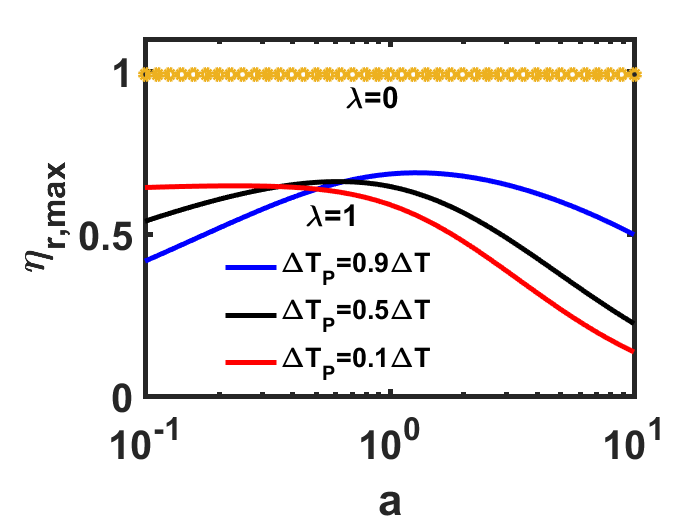}\label{13b}}
		\quad
		\caption{Effect of asymmetric $\gamma^{el}$: For different values of $\lambda$, variation of (a) $J^Q_{C,max}$ and (b) $\eta_{r,max}$ are plotted in semi-log graphs as a function of the multiplication factor ($a$) which scales $\gamma_C$ and $\gamma_H$ to $a\gamma_C$ and $\gamma_H/a$, respectively. Although $J^Q_{C,max}$ is nearly symmetric with respect to $a$, $\eta_{r,max}$ is purely asymmetric and relatively improves for $a>1$ when $T_P \approx T_C$ and for $a<1$ when $T_P \approx T_H$. The intermediate state of $T_P=0.5(T_C+T_H)$ also performs better when $a<1$.}
		\label{gamma}
	\end{figure}

	This section outlines the criteria for maximizing $J_{C,max}^Q$ and $\eta_{r,max}$ by tuning the coupling rates ($\gamma^{ph}$, $\gamma^{el}$) with different reservoirs based on the role played by the phonon bath. First, we investigate the effect of dot-to-bath coupling $\gamma^{ph}$ in manipulating the CP and COP. Primarily, when $\lambda=0$, the dot phonons equilibrate via the phonon bath and the dot-to-bath phonon flow ceases. As a result, both the CP and COP remain unaltered with respect to $\gamma^{ph}$. However, switching on a finite $\lambda$ drives the dot phonon out of equilibrium and kick-start $J_P^{Qph}$. The temperature and the average phonon population of the dot ($\langle N_{ph} \rangle$) are then primarily determined by the applied bias and the temperature of the electronic reservoirs. Here, we consider three different thermal states of the bath by varying its temperature ($T_P=T-\Delta T_P$) as $\Delta T_P=0.9\Delta T$, $0.5\Delta T$, and $0.1\Delta T$, respectively.
	In the first case, when $T_P$ tends to $T_C$, $\langle N_{ph} \rangle$ exceeds the equilibrium bath phonon population and the dot relaxes phonon through the bath throughout the allowed range of $V_{app}$. In contrast, when $\Delta T_P=0.1\Delta T$, i.e. $T_P$ approaches $T_H$, the bath injects phonons into the dot except for the high bias range. For the intermediate case of $\Delta T_P=0.5\Delta T$, the bath, depending upon the temperature of the dot, can switch its role within the bias window.\\ 
	\indent Figures \ref{beta}(a) and \ref{beta}(b), respectively, exhibit the variation of $J_{C,max}^Q$ and $\eta_{r,max}$ with respect to $\gamma^{ph}$ at $\lambda=1$ for the three cases stated above. These plots suggest that for $\Delta T_P=0.9\Delta T$, both $J_{C,max}^Q$ and $\eta_{r,max}$ improve with increasing $\gamma^{ph}$, however, the picture becomes exactly opposite for $\Delta T_P=0.1\Delta T$ where a low $\gamma^{ph}$ is preferred for better performance. The intermediate state does not seem to have much effect with the variation of $\gamma^{ph}$. To explain such trends, we first note that any change in $J_P^{Qph}$ due to a variation of $\gamma^{ph}$ at a fixed bias is primarily balanced by the heat currents flowing through the other reservoirs. In particular, this model shows that a change in the heat flow through the bath in either direction is primarily balanced by the heat current flowing through the cold terminal. This phenomenon gets clearly reflected from Fig.~\ref{beta}(a) where $J_{C,max}^Q$ rises with increasing $\gamma^{ph}$ for $\Delta T_P=0.9\Delta T$ since the excess $J_P^{Qph}$ drawn from the dot is supplied by the cold terminal. In the opposite case of $\Delta T_P=0.1\Delta T$, when the bath injects phonon into the dot, a large $\gamma^{ph}$ suppresses the heat coming out of the cold electronic terminal. A similar trend is also observed for $\eta_{r,max}$ as depicted in Fig.~\ref{beta}(b). The intermediate state of the bath for $\Delta T_P=0.5\Delta T$ does not have pronounced effect, since, in this condition, the temperature of the bath inches close to the that of the dot and the dot-to-bath phonon current becomes weak. This study suggests that in order to optimize the peak performance, the design of phonon coupling should be synchronized with the thermal engineering of the third terminal.\\
	\indent Finally, we study the effect of asymmetry in electronic couplings in optimizing thermoelectric refrigeration. This study is essential since in practical nanoscale multi-terminal setups, couplings between the conducting channel and the different reservoirs are often asymmetric in nature \cite{Leroy,leijnse}. Incorporating such asymmetry in our simulation, in due course, we introduce a multiplying factor $a$ to scale $\gamma^{el}_H$ to $\gamma^{el}_H/a$ and $\gamma^{el}_C$ to $a\gamma^{el}_C$ such that their product remains constant. Since $\gamma^{ph}<<\gamma^{el}$, such an asymmetry in $\gamma^{ph}$ does not make any notable difference and therefore is not ideally considered. 
	Figures \ref{gamma}(a) and \ref{gamma}(b), respectively, capture that both $J_{C,max}^Q$ and $\eta_{r,max}$ maximize around $a=1$ and decline as $a$ is increased or decreased further. However, there is a clear contrast between these two trends which needs an in-depth analysis. \\
	\indent From elementary network theory, we know that for asymmetric $\gamma^{el}$, the conductance of the channel becomes proportional to an effective $\gamma^{el}$ which is even lesser than the weaker $\gamma^{el}$. Hence, the charge current and CP should maximize at $a=1$ and fall down as $a$ is increased or decreased further. However, a close observation of Fig.~\ref{gamma}(a) reveals that $J_{C,max}^Q$ is slightly asymmetric around $a=1$ and maximizes when $\gamma_H^{el}$ is marginally higher than $\gamma_C^{el}$. This effect arises due to the finite Coulomb interaction inherently present in the system. From our previous observation, it is evident that for a $n$-type system, refrigeration happens at $V_{app}>0$. In this case, due to the finite Coulomb interaction, both $I$ and $J_C^Q$ become relatively higher for $a<1$. This can be established by simple analytics described in standard quantum transport literature \cite{DattaQT,dattaLNE}. However, since the margin of asymmetry in $J_{C,max}^Q$ with respect to $a$ is insignificant and invariant of $T_P$, the result shown in Fig.~\ref{gamma}(a) does not hold much significance in the present context of the study. \\
	\indent On the other hand, the variation of $\eta_{r,max}$ with $a$ as shown in Fig.~\ref{gamma}(b), is purely asymmetric in nature and depicts some interesting patterns as we vary $T_P$ within the range $(T_H,T_C)$. To explain this, we first note that for an asymmetric coupling at a finite $\lambda$, the temperature of the dot is essentially governed by the contact to which it is strongly coupled. For example, a stronger coupling with the cold contact pulls down the dot temperature close to $T_C$. Moreover, the temperature gradient between the bath and the dot determines the magnitude and direction of $J_P^{Qph}$. Initially, at zero $\lambda$, $\eta_{r,max}$ remains invariant with the asymmetry parameter due to the absence of finite $J_P^{Qph}$. However, at finite values of $\lambda$ (here, $\lambda=1$), the peak cooling efficiency depends closely on the temperature of bath. In the case of $\Delta T_P=0.9\Delta T$, a strong coupling with the cold contact (i.e $a>1$) is necessary to optimize $\eta_{r,max}$ since in this limit less heat is dragged out from the bath. In contrast, for $\Delta T_P=0.1\Delta T$, a strong coupling with the hot contact (i.e $a<1$) is favorable to accomplish the same job to allow stronger phonon relaxations. 
	This signifies that for a given $J_C^Q$, the requirement of input work at $V_{app}=V_{\eta_{r,max}}$ is relatively less for a stronger coupling with the hot (cold) contact when $T_P$ is closer to $T_H$ ($T_C$). 
	\section{Conclusion}
	\label{conclu}
	This paper presented a three terminal nanoscale refrigeration concept based on a vibron-coupled quantum dot hybrid system. In setting up the simulation framework, we ensured refrigeration to be driven by the electronic bias while the phonon bath was incorporated to capture the heat exchange with the substrate. Our key results evidenced that the electron-phonon coupling, although being apparently detrimental from a general refrigeration perspective, can be engineered to favorably improve the trade-off between the CP and COP. Furthermore, it was manifested that an additional improvement in the trade-off could be facilitated by a high electronic thermal bias. However, such improvements are largely limited by the electron-phonon coupling since it restricts the lowest achievable temperature of the cold end. The other aspect of this study was aimed to explore the impact of phonon bath and the junction coupling rates in optimizing the refrigeration performance. In particular, it was shown that depending on whether the bath extracts or injects phonons into the dot, one should respectively have a strong or weak coupling between the dot and bath. Finally, the study of asymmetric electronic coupling revealed that a stronger coupling is favorable with the contact which is having temperature closer to that of the bath. To put things together, our work proposed key design guidelines for optimized performance and paves the way for a possible experimental realization of vibron-coupled molecular and quantum dot thermoelectric refrigerator.\\
	{\it{Acknowlegements:}} The authors acknowledge funding from Indian Space Research Organization (ISRO) under the ISRO-IIT Bombay Space Technology Cell . This work is also an outcome of the Research and Development work undertaken in the project under the Visvesvaraya PhD Scheme of Ministry of Electronics and Information Technology, Government of India, being implemented by Digital India Corporation (formerly Media Lab Asia).  Funding from the Science and Engineering Board (SERB), Govenment of India under grant file number EMR/2017/002853.
	\appendix
	\section{Derivation of $\eta_{rev}$ for three-terminal setup} \label{app1}
	The total entropy ($\mathcal{S}$) production rate of the system is calculated by adding the individual contributions from each of the three terminals. For a given terminal $i$, the entropy production rate is defined as $\dot{\mathcal{S}_i}=\frac{\dot{Q_i}}{T_i}$, where $\dot{Q_i}$ is the amount of heat current flowing out of that terminal and $T_i$ is the temperature of that terminal. Therefore, for a dissipating bath, $\dot{\mathcal{S}}$ is given by
	\begin{equation}
		\dot{\mathcal{S}}=\frac{\dot{Q_C}}{T_C}-\frac{\dot{Q_H}}{T_H}-\frac{\dot{Q_P}}{T_P}=\frac{J^Q_C}{T_C}-\frac{J^Q_H}{T_H}-\frac{J^{Qph}_P}{T_P},
		\label{entr}
	\end{equation}
	where positive direction is assumed for particle flow from a terminal to the channel. In the reversible limit, the rate of total entropy production becomes zero. Hence, putting $\dot{\mathcal{S}}=0$ in Eq. \eqref{entr} and using the law of energy balance given by Eq. \eqref{heatbalance}, one obtains
	\begin{equation*}
		\frac{J^Q_C}{T_C}-\frac{J^Q_C+W-J_P^{Qph}}{T_H}-\frac{J^{Qph}_P}{T_P}=0.
	\end{equation*}
	After few simple steps, one gets the final expression of $\eta_{rev}$ for a dissipating bath as
	\begin{equation}
		\eta^{diss}_{rev}=\left| \frac{J^Q_C}{W}\right|_{\dot{\mathcal{S}}=0} =\eta_{rev}^{2T}\left[ 1+ \frac{J^{Qph}_P}{W} \left( {\frac{T_H}{T_P}-1} \right)\right].
	\end{equation}
	On the other hand, when the bath injects phonon into the dot, the direction of $J^{Qph}_P$ reverses and both Eq. \eqref{entr} and Eq. \eqref{heatbalance} are modified by reversing the sign of $J^{Qph}_P$. Calculating the COP given by Eq. \eqref{copex1} for $\dot{\mathcal{S}}=0$, one gets
	\begin{equation}
		\eta^{driv}_{rev}=\left| \frac{J^Q_C}{W+J^{Qph}_P}\right|_{\dot{\mathcal{S}}=0} =\eta_{rev}^{2T}\left[ 1- \frac{T_H}{T_P} \left( \frac{1}{\frac{W}{J^{Qph}_P}+1}\right)\right].
	\end{equation}
	It is important to note from the above two expressions that for finite values of $J^{Qph}_P$ and $W$, one obtains the relation $\eta^{dis}_{rev} > \eta_{rev}^{2T} > \eta^{inj}_{rev}$ which indicates that the upper-bound of COP can be efficiently engineered in both ways by proper designing of hybrid multi-terminal setups.
	
	\bibliography{reference}
\end{document}